# Predicting colorectal polyp recurrence using time-to-event analysis of medical records

Lia X. Harrington, PhD[1], Jason W. Wei[1], Arief A. Suriawinata, MD[2],
Todd A. Mackenzie, PhD[1], Saeed Hassanpour, PhD[1]

[1]**Dartmouth College, Hanover, NH, USA**
[2]**Dartmouth-Hitchcock Medical Center, Lebanon, NH, USA**

**Abstract**

*Identifying patient characteristics that influence the rate of colorectal polyp recurrence can provide important insights into which patients are at higher risk for recurrence. We used natural language processing to extract polyp morphological characteristics from 953 polyp-presenting patients' electronic medical records. We used subsequent colonoscopy reports to examine how the time to polyp recurrence (731 patients experienced recurrence) is influenced by these characteristics as well as anthropometric features using Kaplan-Meier curves, Cox proportional hazards modeling, and random survival forest models. We found that the rate of recurrence differed significantly by polyp size, number, and location and patient smoking status. Additionally, right-sided colon polyps increased recurrence risk by 30% compared to left-sided polyps. History of tobacco use increased polyp recurrence risk by 20% compared to never-users. A random survival forest model showed an AUC of 0.65 and identified several other predictive variables, which can inform development of personalized polyp surveillance plans.*

**Introduction**

Colorectal cancer (CRC) is the third most common cancer in the United States, killing over 50,000 individuals each year [1]. CRC often develops from adenomas, polyps arising from glandular epithelial tissue [2]. Accumulation of genetic and epigenetic mutations in these adenomas can result in CRC progression [3]. CRC can also arise from hyperplastic polyps, which previously had been thought of as benign lesions [4]. Surveillance of the colorectal region via colonoscopies has proved extremely successful in preventing CRC, as polyps can be removed before they turn cancerous [5]. Depending on the size, number, and histology of polyps, various surveillance plans are routinely recommended [6].

Previous studies have examined time-to-death for patients with CRC [7,8]. However, the relationship between precancerous polyps and time to subsequent polyp development in the context of multiple relevant risk factors has not been comprehensively modeled, despite the fact that polyp recurrence is quite common. The associated rates of recurrence within the 1-, 3-, and 5-year surveillance periods are 10.9%, 38.2%, and 52.5%, respectively [9]. While polyp recurrence is common, the fact that patients experience it at different rates, including some who do not develop subsequent polyps at all, suggests the existence of risk factors that influence susceptibility to polyp recurrence. This differential susceptibility motivates investigation into how individual traits and characteristics influence the rate of polyp recurrence. Variables linked previously to greater chances of recurrent polyps include the number of polyps at baseline [10], polyp size [11], polyp location [12], and gender [13].

Time-to-event analysis, also known as survival analysis, is one useful approach to understanding the influence of various factors on the amount of time that elapses before the occurrence of a certain event (polyp recurrence, in this case), and can aid in determining which factors drive those differences [14]. Although time-to-event analysis has been performed previously in the context of colorectal polyp recurrence [9,15], patient-level predictive modeling was not performed, nor were specific risk factors described. Another study [10] performed basic predictive modeling with physician-curated records, but did not consider free-text medical records and colonoscopy reports, which may contain a wealth of additional information that could be potentially useful in modeling the factors that drive polyp recurrence. In this study, we focused on identifying variables that influence the risk of polyp recurrence by using demographic and clinical information obtained from electronic medical records (EMR), as well as polyp characteristics automatically extracted from colonoscopy records, to develop a natural language processing (NLP) pipeline. We expect that the integration of our findings into CRC surveillance programs could potentially improve risk assessment and follow-up recommendations for patients.

**Methods**

*Patient cohort selection*

The colonoscopy information of 4,273 randomly selected patients who underwent colonoscopy from 2011 to 2017 was obtained from Dartmouth-Hitchcock Medical Center (DHMC; Lebanon, NH), a tertiary academic care center. Use of human subject data was approved by Dartmouth Institutional Review Board with a waiver of informed consent. Of note, the patient records before this period were not available for this study due to the installation of a new EMR system at DHMC in 2011. This data was filtered using the following inclusion/exclusion criteria for each patient:

1. At least one record of a polyp must exist,
2. Records must display no evidence of colitis or Crohn's disease (patients who are usually under intense surveillance),
3. Records of two or more colonoscopies must exist, separated by at least 6 months (records that occurred less than 14 days after another visit were assumed faulty due to poor colon preparation).

The first record containing polyp information was considered the baseline time point for that patient ($t = 0$) and was paired with the first subsequent record showing recurrence (if the patient was recurrent) or the last available record (if the patient was non-recurrent). These selection criteria resulted in 953 patient record pairs included for further analysis, while 3,320 patients (77.7% of the original dataset) were excluded for failure to meet these criteria. Of these remaining pairs, we observed 731 polyp recurrence events and 222 non-recurrences (76.7% recurrence rate over 6.25 years). 712 recurrences and 214 non-recurrences had completely non-missing data for all variables considered in this study and were included in the final Cox proportional hazards and random survival forest models; hence, 27 patients were excluded due to missing data. All information except sample time and recurrence status was propagated to the subsequent records from the first, allowing us to predict time to recurrence from information available as of the first colonoscopy only.

*Extraction of polyp characteristics from EMR*

To obtain information about colorectal polyp characteristics, we developed an in-house natural language processing (NLP) information extraction pipeline written in Python 3.6 (Python Software Foundation, Beaverton, OR). This pipeline leveraged the NLTK python library [16] to mine colonoscopy reports and extract information relating to polyp sizes, numbers, and locations, as these polyp features have previously been found to be important in the development of colorectal cancer [10–12]. Our NLP pipeline allowed us to completely automate the information extraction process, from retrieving EMRs to finalizing the variables prepared from text, without manual curation.

First, we scanned for the presence of known colonic locations from a controlled vocabulary used by DHMC that included the following locations: transverse, sigmoid, ileum cecum, anus, ascending, descending, hepatic, rectum, ileocecal, and splenic. Additionally, both textual and digit representations of numbers were detected using a dictionary lookup that converts both representations to floating point variables. Whenever a number was identified, the script determined whether it corresponded to a size or a quantity of polyps by identifying whether units ("mm" or "cm") were present alongside or closely after the number.

We reconciled the correspondence between polyps and their features by creating lists of polyp sizes and locations in the order in which they were found by the parser. Both polyp location and sizes for each polyp were reported in this same order for the total number of polyps in each colonoscopy record. Finally, we aggregated polyp information for each visit by averaging polyp sizes and incrementing a master list of locations in which polyps were found during that visit. Because sizes were often reported as a range (minimum to maximum size for each polyp), both size bounds were parsed (one value was used for both bounds if only a single size value was provided).

*Incorporation of patient demographics and clinical data*

Demographics and clinical data were queried from the DHMC EMR (Epic, Verona, Wisconsin) and were merged with the extracted polyp information for our analysis. This data includes anthropometrics such as gender, age, body mass index (BMI), smoking status, marital status, race, and ethnicity. Records were excluded from the final Cox proportional hazards analysis in cases of missing data. Continuous variables such as age and BMI were factorized by using the median to demarcate lower and higher bins. For polyp location, if most polyps in a given visit were localized in the left colon, it would receive the "left" designation, and vice versa. The designation of "other" was conferred in cases where polyps did not localize to either side more.

*Model Development*

Kaplan-Meier curves were generated and log-rank tests were performed to determine the significance of each variable in influencing time to recurrence. Variables with log-rank *p < 0.2* were used in generating a Cox proportional hazards model using the survival package in R [17]. A forest plot was used to visualize the resulting risk ratios with the ggForests package [18]. A random survival forest model was generated using the randomForestSRC package [19]. As random forest models are adept at handling many potentially redundant features, unlike Cox proportional hazards models, all variables in the dataset were used in the survival random forest model [20]. Because traditional Kaplan-Meier analyses depend on discretized factors, we created binned variables for our analysis using median and tertile partitions. Thus, binary (greater or less than the median) and/or factorized (tertile) versions of continuous variables such as age, height, number of polyps, and polyp size were created.

**Results**

We plotted Kaplan-Meier time-to-event curves to assess whether various EMR features influence time to polyp recurrence in our cohort, shown with a common censor time of 95% of the total time period 21. These factors include age, gender, BMI, height, weight, smoking status, smoking frequency, race, ethnicity, marital status, polyp count, polyp location, and polyp size. Of these features, size of polyps (p = 0, Fig. 1) significantly altered time to recurrence: with increasing size, the time to recurrence shortens.

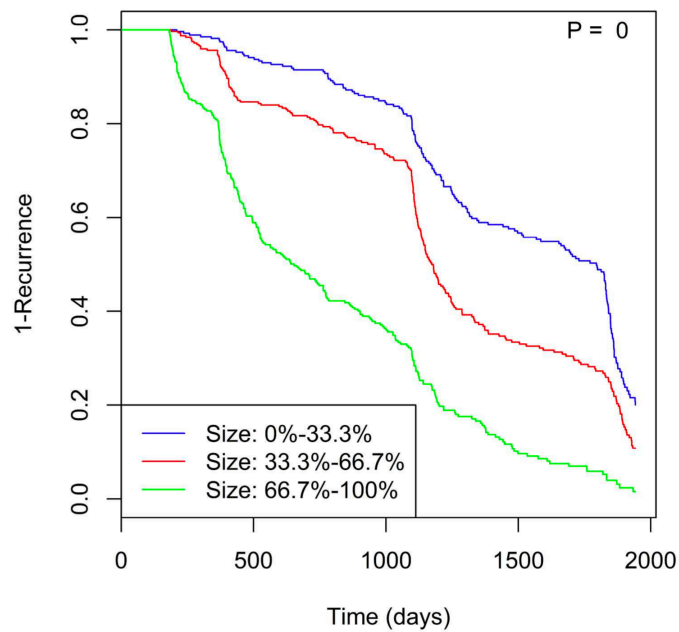

**Figure 1.** The time to polyp recurrence becomes shorter with increasing polyp size at the baseline screening.

Furthermore, the number of polyps (*p* = 0, Fig. 2) likewise shows that having more polyps shortens the recurrence timeframe. Tobacco use (*p* = 0.056, Fig. 3) also may hasten time to recurrence. Here, the "Never" class indicates patients who have never used tobacco, while the "Used" class consists of those who have (including current and former regular smokers). Those who used tobacco showed significantly reduced time to polyp recurrence. Finally, polyp location played a significant role ($p$ = 0.032, Fig. 4), with right colon localization coinciding with a significantly increased polyp recurrence rate compared to the left or other regions.

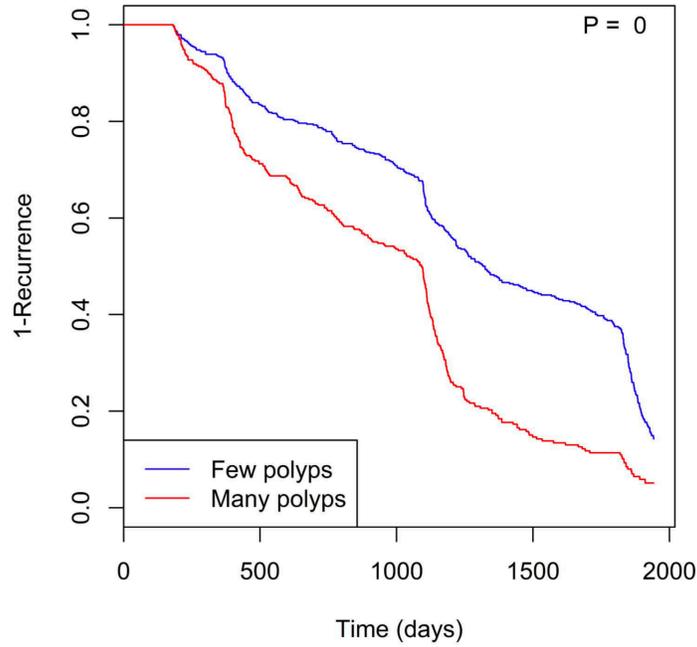

**Figure 2.** Having more than the median number of polyps (i.e., two) at the first colonoscopy coincides with more rapid polyp recurrence.

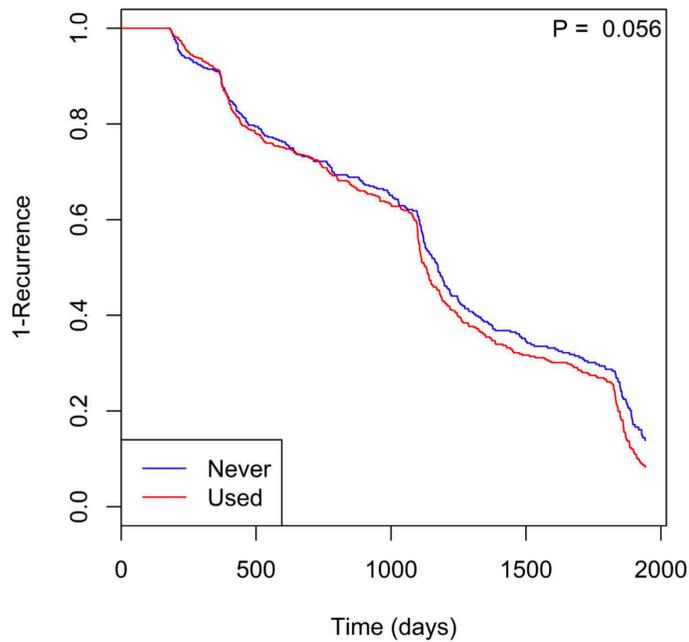

**Figure 3.** Tobacco use (current or past) may reduce time to polyp recurrence.

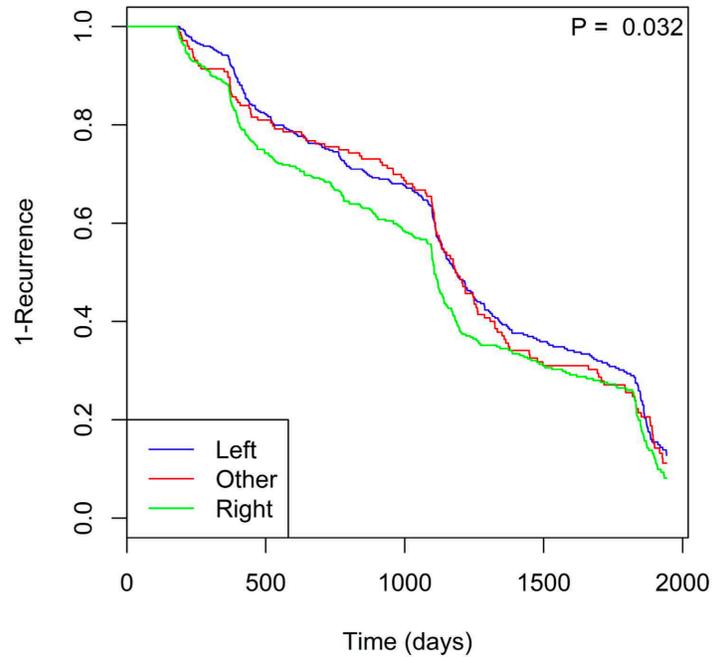

**Figure 4.** Polyp time to recurrence is faster when polyps localize to the right colon.

In additional analyses, we observed that other factors such as BMI and gender, which were previously considered to influence polyp recurrence or colorectal cancer, showed weak or statistically insignificant associations to polyp recurrence. Similarly, all other factors included in our analysis did not show significant associations with time to recurrence.

All informative features according to this Kaplan-Meier time-to-event analysis with $p < 0.2$ (polyp number, polyp size, tobacco use, colon location, BMI) were used to create a Cox proportional hazards model for polyp recurrence. Wherever continuous versions of the same (discretized) variables existed, they were used in place of the discretized variants shown in the Kaplan-Meier analysis; namely, polyp number, size, and BMI. The model is overall significantly predictive ($\chi^2(6) = 228$, $p = 0$), and the risk ratios with 95% confidence intervals for each risk are provided in Fig. 5. The recurrence risk ratio (RR) increases with increasing number of polyps (RR = 1.1; 95% CI: 1.07-1.1) and maximum polyp size (RR = 1.1; 95% CI: 1.07-1.1). Use of tobacco increases polyp recurrence risk by 1.2 (95% CI: 1.03-1.4) times that of never-users. Finally, having polyps primarily in the right colon confers 1.2 (95% CI: 1.07-1.5) times the risk of polyp recurrence compared to the left colon.

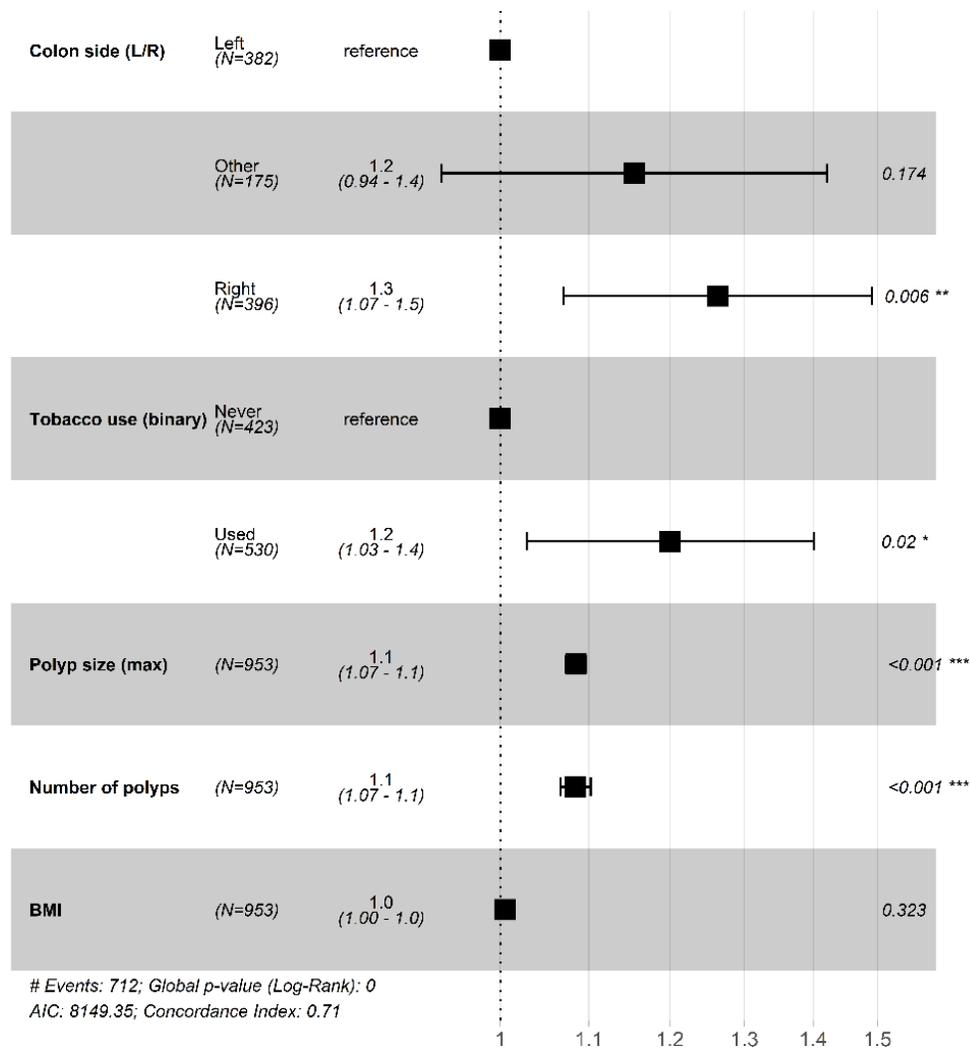

**Figure 5.** Forest plot of risk ratios from our Cox proportional hazards model for polyp recurrence. The dashed line represents the reference (baseline) risk ratio of 1.0 (no increase or decrease in risk).

We also produced a random survival forest model to predict time to polyp recurrence. A random survival forest model is a modification of traditional random forest models using a set of survival-tree-specific splitting functions (based on default log rank splitting rule [22]), prediction objectives (cumulative hazard function), and evaluation criteria (Harrell's Concordance error rate for out-of-bag/OOB error estimation) [19]. The model was generated with 1,000 trees and all variables mentioned above, including both continuous and discretized versions of the same variables when applicable, for a total of 37 predictors. The random survival forest model produced an out-of-bag error rate of 28.48% for the whole time period, as well as an area under the curve (AUC) score of 0.65 (Fig. 6) when computed at a prediction timepoint of 1,500 days. The prediction timepoint of 1,500 days was chosen as the ground truth for the AUC calculation as there is fairly clear separation in the plotted survival curves around this period. The most important features reported by the model are shown in Fig. 7.

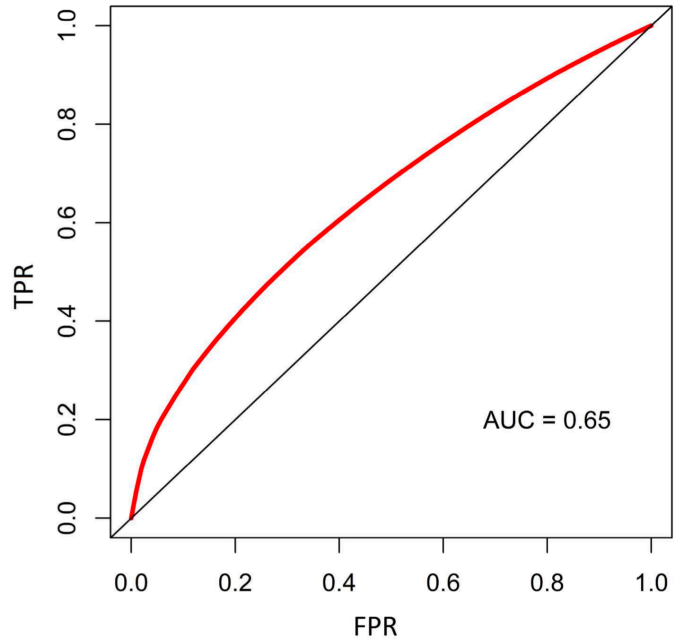

**Figure 6.** The receiver operator characteristic (ROC) curve for the random survival forest model predictions at 1,500 days from initial polyp detection shows an AUC of 0.65. True positive rates (TPR) are shown on the y-axis and false positive rates (FPR) on the x-axis.

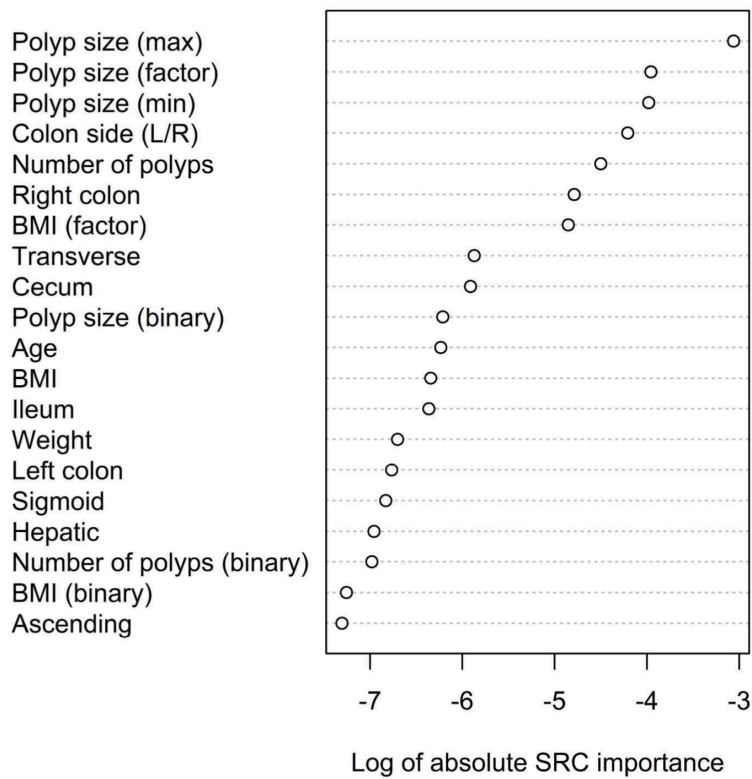

**Figure 7.** Variable importance scores shown are the log10 of the absolute value of the random survival forest model importance weights. Polyp features (size, number, location) and BMI dominate the listing.

**Discussion**

We analyzed EMR records of polyp information and applied time-to-event analysis to characterize polyp recurrence. Kaplan-Meier curves, Cox proportional hazards models, and random survival forest models were explored in order to demonstrate the effects of polyp characteristics, patient demographics, and clinical information on the rate of polyp recurrence. We found features associated with decreased time to polyp recurrence that may prove to be important in tailoring patient surveillance plans based on patient health data and initial colonoscopy results.

In particular, polyp size and number were found to be important for increasing the risk of polyp recurrence. This finding is consistent with previous studies, including the current colonoscopy surveillance guidelines [6] which suggest that greater size [23] and increased number of polyps [24] correspond to polyps at a heightened risk of developing into cancer. One possible explanation for this association is that the probability of deleterious mutations arising may increase with polyp size. Another potential hypothesis is that the increased volume of polyp tissue serves as a biomarker for high underlying mutation burden in the surrounding tissue. In addition to polyp size and number, tobacco use was also associated with differential time to polyp recurrence. Tobacco use has previously been linked to development of hyperplastic polyps [25], despite being found to be of only marginal significance in a previous study [10]. Furthermore, smoking has been linked to both the greater presence of distal versus proximal adenomas as well as to increasing multiple versus single adenomas [26]. Interestingly, the latter association may to some extent be driving the increased polyp number mentioned previously.

Notably, the general colonic location of the polyps recovered in the initial colonoscopy was found to be a significant predictor of polyp recurrence, with polyps located in the right colon predisposing patients to higher recurrence risk. The histological features of the two sides are known to differ at a molecular level [27], and there are previous reports of poorer prognosis for patients diagnosed with right-colon CRC [28], which the authors attribute to potential increased mutation burden in the more ileocecal-proximal colon (classically, the right colon). It can be inferred that a higher risk of polyp recurrence in the region is consistent with the comparative aggressiveness of right-sided CRC. The increased recurrence rate in the right colon, however, might be confounded by left-sided polyps being easier to detect due to their polypoid morphology [29].

To enable group comparisons in the Kaplan-Meier analyses, some categorical variables were generated from continuous variables (e.g., BMI, polyp size, and number). However, the reduction in dynamic range when discretizing continuous variables may lead to inferior performance in models capable of utilizing continuous variables, such as the Cox proportional hazards model and the random survival forest model. We found that swapping in the continuous versions of the discretized variables that were shown to be significant with the survival curve analysis produced more significant coefficients in the Cox proportional hazards model. Hence, consistent with intuition, the continuous versions of variables considered in this study tend to be more informative than their discretized counterparts.

The random survival forest model AUC score of 0.65 and the reasonably low out-of-bag (OOB) error rate of 28.48% are indicative of promising predictive model performance. Comparatively, an AUC score of 0.5 denotes performance of a random model, much like an OOB error rate of 50% in the case of random survival forest models [19]. Unlike AUC calculation, the OOB error rate in random survival forest models is derived from Harrell's concordance index [29], and thus is not dependent on specifying a prediction timepoint [19]. Interestingly, BMI performed as a strong predictor in the random survival forest model in comparison to the Cox proportional hazards model and survival curve analysis. This may be due to the complementary role of BMI information in relation to other covariates which were held out of the simpler models but included in the random survival forest model.

Overall, our analysis shows time-to-event (or "survival") analysis is a powerful technique for elucidating the factors that drive polyp recurrence. Notably, use of NLP afforded extraction of polyp size, number, and location from colonoscopy records and these were critical in differentiating and predicting time to recurrence. Because polyp recurrence is an important risk factor for the emergence of colorectal cancer, this analysis may have implications for CRC diagnosis as well. Some of the most significantly associated factors discovered by this analysis are consistent with previous work, but are further extended in this paper to introduce clinically relevant risk ratios and predictive models. As a result, this analysis can be useful for establishing a patient-specific risk of polyp recurrence. Additionally, using the proposed predictive machine learning model, we can estimate time to recurrence based on patient clinical data and polyp characteristics available from the initial colonoscopy. Such a model can usher in a precision-medicine-based approach for personalizing CRC surveillance plans.

Importantly, our time-to-event analysis pipeline with EMR data extends to predicting other outcomes such as CRC itself and even other cancers and cancer precursors. For example, the risk of breast cancer is known to increase by at least four-fold in the presence of high-risk breast lesions such as atypical ductal hyperplasia and atypical lobular hyperplasia [31]. It is feasible to extend the current pipeline to prediction of time to developing new cases of high-risk lesions given patient factors such as age, hormonal molecular subtype, and other characteristics derived from patient medical records. One strong advantage of using NLP to analyze EMR is that this allows extraction of any predictors contained in these records as opposed to being limited to only curated, structured data.

There are some limitations in the presented study. First, edge cases are possible in our dataset where the first record containing polyp information is not the first true incident. In these cases, this information would be recorded in other EMR systems at other institutions. To mitigate this risk, we plan to collaborate with a state-level colonoscopy data registry for a more comprehensive data collection and to makes these occurrences less likely. In addition, there might be cases where a recurrent polyp was not truly new but was instead an existing polyp that was missed in the previous colonoscopy. In the current study, we cannot separate these two cases, but we do require that all patients have a baseline polyp in our patient inclusion/exclusion criteria.

Refinement of our models with data from more patients outside of a single medical institution may further improve predictive performance as well as generalizability. This may be particularly salient in the present case where over 76% of patients experienced polyp recurrence in the 6.25 years of data available, a higher number than previously reported elsewhere [9]. Practically, for the collection of this particular type of data in the future, the inclusion of structured polyp information fields in medical records would dramatically reduce errors associated with NLP of unstructured text. Ultimately, it is our goal to enable the optimization of colonoscopy administration through a paradigm in which the frequency of each patient's follow-ups is determined according to that patient's predicted time to polyp recurrence. Achieving this goal, which we will pursue in future work, may allow physicians to spare low-risk patients unneeded colonoscopies while more aggressively monitoring patients at greater risk of rapid polyp recurrence.

**Conclusion**

In closing, our study evaluated how morphological characteristics of colorectal polyps influenced the rate of polyp recurrence. We found that polyp size, number, location, and patient smoking status correlated significantly with the recurrence rate. Moreover, colon polyps on the right-side increase recurrence risk by 30% compared to polyps on the left-side, and a history of tobacco use increased recurrence risk by 20% compared with never-users. Finally, we trained a random survival forest model for predicting survival that achieved an AUC of 0.65 and identified other predictive characteristics that could be helpful for developing personalized polyp surveillance plans.

**Acknowledgements**

The authors would like to thank Lamar Moss for his feedback on this paper. This work was supported in part by the Burroughs Wellcome Big Data in the Life Sciences Fellowship and grants from the National Institutes of Health (R01LM012837 and P20GM104416).